\documentclass{article}
\usepackage{flushend}
\usepackage{spconf}

\usepackage{url}
\usepackage{enumitem}
\usepackage{stfloats}
\usepackage{romannum}
\usepackage{etoolbox}
\robustify\bfseries
\usepackage{graphicx}
\usepackage{amsmath,amssymb}
\usepackage[ruled,linesnumbered]{algorithm2e}
\SetAlCapNameFnt{\footnotesize}
\SetAlCapFnt{\footnotesize}
\usepackage{tabularx}
\usepackage{makecell}
\usepackage{booktabs}
\usepackage{multirow}
\usepackage{arydshln}
\usepackage{color,soul}
\usepackage{cite}
\usepackage{balance}
\usepackage{hyperref}
\usepackage{siunitx}
\usepackage{caption}
\def\H{{\mathsf H}}
\def\T{{\mathsf T}}
\def\CC{{\mathbb C}}

\makeatletter
\newcommand*{\bigs}[1]{{\hbox{$\left#1\vbox to9\p@{}\right.\n@space$}}}
\makeatother
\title{Neural Forward Filtering for Speaker-Image Separation}
\name{Jingqi Sun, Shulin He, Ruizhe Pang, and Zhong-Qiu Wang
}
\address{Department of Computer Science and Engineering\\
Southern University of Science and Technology, Shenzhen, China\\
\texttt{\small jingqi.sun@outlook.com, wang.zhongqiu41@gmail.com}
}
\makeatletter
\renewcommand\normalsize{%
   \@setfontsize\normalsize{9pt}{10.5pt}} 
\makeatother
\begin{document}

\maketitle

\begin{abstract}%
We address monaural multi-speaker-image separation in reverberant conditions, aiming at separating mixed speakers but preserving the reverberation of each speaker.
A straightforward approach for this task is to directly train end-to-end DNN systems to predict the reverberant speech of each speaker based on the input mixture.
Although effective, this approach does not explicitly exploit the physical constraint that reverberant speech can be reproduced by convolving the direct-path signal with a linear filter.
To address this, we propose CxNet, a two-DNN system with a neural forward filtering module in between.
The first DNN is trained to jointly predict the direct-path signal and reverberant speech.
Based on the direct-path estimate, the neural forward filtering module estimates the linear filter, and the estimated filter is then convolved with the direct-path estimate to obtain another estimate of reverberant speech, which is utilized as a discriminative feature to help the second DNN better estimate the reverberant speech.
By explicitly modeling the linear filter, CxNet could leverage the physical constraint between the direct-path signal and reverberant speech to capture crucial information about reverberation tails.
Evaluation results on the SMS-WSJ dataset show the effectiveness of the proposed algorithms.
\end{abstract}
\begin{keywords}
speaker-image separation, neural forward filtering, reverberation tail modeling, deep learning
\end{keywords}
\section{introduction}%
\label{sec:introduction}

In the past decade, deep learning has dramatically advanced speaker separation in reverberant conditions \cite{2018_WDL_SSP_overview, 2016_R.Hershey_SS_deep_clustering, 2017_Kolbak_SS_pit}.
Many studies target at not only separating mixed speakers but also suppressing the reverberation of each speaker, as reverberation is harmful for many downstream tasks such as 
robust automatic speech recognition (ASR) \cite{2018_Barker_ASR_CHiME5, 2020_Watanabe_ASR_CHiME6, 2020_Haeb-Umbach_ASR_overview}, speaker recognition \cite{2020_Taherian_SR_time_frequency_masking}, and diarization \cite{2022_Park_SD_overview}.  
Differently, some other studies
aim at separating mixed speakers but preserving the reverberation of each speaker \cite{2021_wang_SIS_sequential}, as reverberation carries essential information about the acoustic environment \cite{2025_nakata_reverbmiipher}.
By preserving reverberation, we can enable operations such as source volume adjustment, reverberation level control, and source replacement, all of which are very useful features in applications such as augmented reality \cite{2022_Gupta_AR_overview} and audio post-production \cite{2022_Petermann_PP_DnR, 2022_Petermann_PP_benchmark_several_models}.
We refer to this task as \textit{speaker-image separation} and the former as \textit{speaker separation}, and this paper deals with speaker-image separation.

In speaker-image separation, although reverberation is not required to be removed, preserving the reverberation of each speaker is still a challenging task, as late reverberation itself is too weak to be separated and reconstructed.
Unlike direct-path signals, which exhibit clear spectro-temporal patterns, late reverberation arrives at the microphone from multiple directions and can be considered a diffuse source \cite{1993_polack_RIR_billiards}.
It often lacks distinct spectro-temporal cues that could be exploited for separation, particularly in time-frequency (T-F) units dominated by the reverberation of different speakers.
This problem poses difficulties for purely supervised learning based approaches for speaker-image separation, where DNN models are trained to directly predict the reverberant speech of each speaker in an end-to-end fashion \cite{2010_mandel_SSP_reverb_evaluating} overlooking the physical filtering relation between direct-path and reverberant signals.

To address this problem, our key idea is, besides estimating reverberant speech, additionally estimating the direct-path signal of each speaker and the relative transfer function (RTF) relating the direct-path signal to reverberant speech.
Once they can be accurately estimated, their linear-convolution results can be utilized as an estimate of the reverberant speech, which could be leveraged in turn as a discriminative input feature to improve supervised speaker-image separation.
Building on this idea, our proposed system, \textbf{CxNet}, employs a sandwich design, where a linear convolutive prediction module lies between two DNN modules.
The first DNN is trained in a supervised way to simultaneously estimate the direct-path signal and reverberant speech of each speaker.
With the estimated direct-path signal, we leverage a neural forward filtering algorithm named forward convolutive prediction (FCP) \cite{2021_Wang_DEREVERB_FCP_journal} and its variants (newly-proposed in this paper) to estimate the RTF, which is then convolved with the direct-path estimate to obtain another estimate of reverberant speech.
Next, the second DNN takes as input features (a) the estimated direct-path signal and reverberant speech by the first DNN; (b) the estimated reverberant speech by the FCP module; and (c) the original
mixture, and is trained in a supervised way to further estimate the reverberant speech of the target speaker.
Notice that CxNet explicitly models the convolutional relationship between direct-path signals and their reverberant images, enforcing a physical constraint derived from room acoustics.
Evaluation results on the public SMS-WSJ dataset \cite{2019_Drude_SSP_SMS_WSJ} show the effectiveness of the proposed algorithms.
The contributions of this paper can be summarized as follows:
\begin{itemize}[leftmargin=*,noitemsep,topsep=0pt]
\item We propose a neural forward filtering approach for speaker-image separation, achieving clear performance gains.
\item We propose a joint prediction framework that simultaneously predicts the anechoic signal and reverberant speech of each speaker, resulting in better estimation of reverberant speech.
\item We propose an extension of FCP with energy-sorted source update (FCP-ESSU), which can better estimate RTFs.
\end{itemize}

\section{Proposed System}
\label{sec:systems_overview}

Given a mixture of $C$ speakers recorded in noisy-reverberant conditions by a single microphone, the physical model in the short-time Fourier transform (STFT) \cite{1983_Nawab_stft} domain can be formulated as Eq.~(\ref{eq:phy_model}), where at time $t$ and frequency bin $f$, $Y(t, f)$, $N(t, f)$, and $X(c, t, f)$, $S(c, t, f)$, and $H(c, t, f)\in \CC$ are respectively the STFT coefficients of the mixture, reverberant noise, and reverberant speech, direct-path signal, and non-direct signals of speaker $c$.
\begin{align}
Y(t,f) & = \sum\nolimits_{c = 1}^{C}X(c, t, f) + N(t, f) \nonumber \\
& = \sum\nolimits_{c = 1}^{C}\Big(S(c, t, f) + H(c, t, f)\Big) + N(t, f). \label{eq:phy_model}
\end{align}
Following \cite{2019_Drude_SSP_SMS_WSJ}, we assume that the noise is weak.
In the rest of this paper, we omit indices $c$, $t$ and $f$ when denoting spectrograms.
Based on the input mixture $Y$, we aim at recovering the reverberant speaker images $X$ (i.e., reverberant speech of each speaker).

\begin{figure}
\centering
\includegraphics[width=8.5cm]{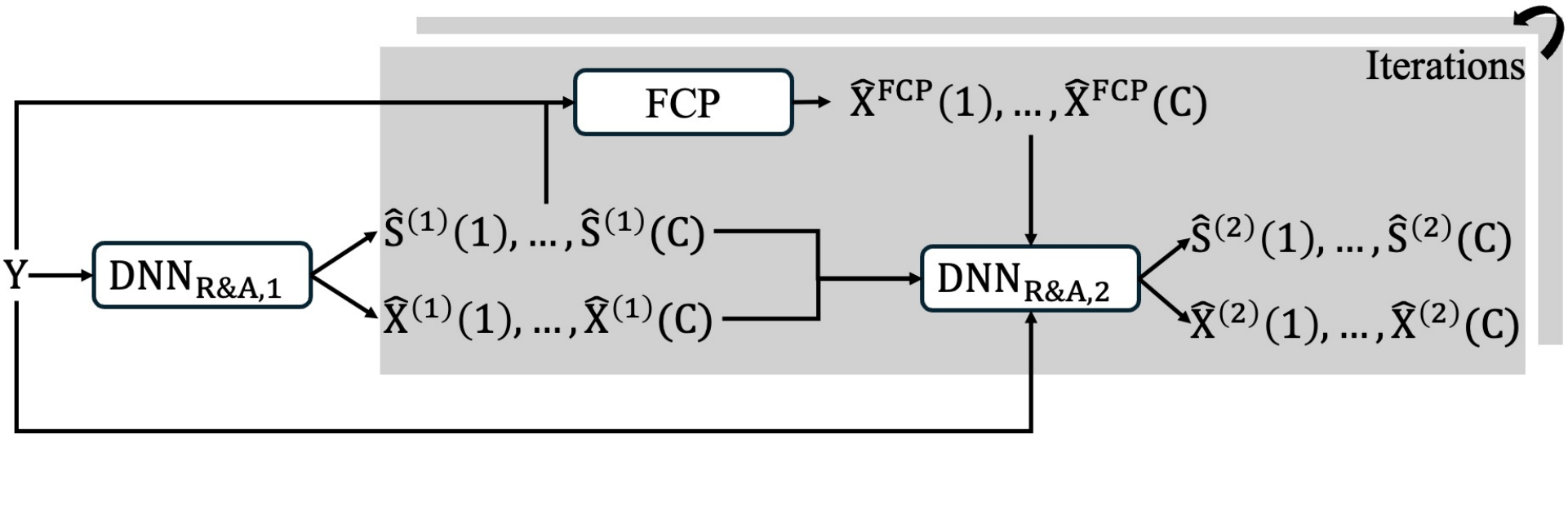}\vspace{-0.6cm}
\caption{
Illustration of CxNet, with structure DNN$_{\text{R\&A,1}}$+FCP+DNN$_{\text{R\&A,2}}$.
}
\label{figure:system_overview}
\vspace{-0.6cm}
\end{figure}

\vspace{-0.2cm}
\subsection{Forward Filtering for Speaker-Image Separation}

Fig.~\ref{figure:system_overview} illustrates our proposed system, CxNet, which consists of two DNN modules with a forward filtering module in between.

The first DNN, denoted as DNN$_{\text{R\&A,1}}$, takes the multi-speaker mixture $Y$ as input, and is trained to produce, for each speaker $c$, an estimate of the direct-path signal, $\hat{S}^{(1)}(c)$, and an estimate of the reverberant speech, $\hat{X}^{(1)}(c)$.
The subscripts ``R'' and ``A'' in DNN$_{\text{R\&A,1}}$ mean that we predict both \textbf{R}everberant and \textbf{A}nechoic signals.

Next, for each speaker $c$, the direct-path estimate $\hat{S}^{(1)}(c)$ is linearly filtered by an estimated RTF (relating the direct-path signal to reverberant speech) produced by a neural forward filtering algorithm named FCP \cite{2021_Wang_DEREVERB_FCP_journal}, which will be described later in Section \ref{FCP_description}.
The output $\hat{X}^{\text{FCP}}(c)$ can be viewed as a physically-constrained estimate of reverberant speech, as it is produced by linear filtering.

Finally, the mixture $Y$ and the estimates $\hat{X}^{(1)}$, $\hat{S}^{(1)}$, and $\hat{X}^{\text{FCP}}$
are combined and used as input for the second DNN, denoted as DNN$_{\text{R\&A,2}}$, to refine the estimation of the speaker image and the direct-path signal (for each speaker $c$, we denote the estimates as $\hat{X}^{(2)}(c)$ and $\hat{S}^{(2)}(c)$).
This way, $\hat{X}^{(2)}$ could benefit from the strong modeling capability of the DNN and at the same time incorporating the physical constraints imposed by the forward filtering module.

At run time, the second DNN can be executed iteratively.
Each iteration benefits from progressively improved estimates of the direct-path signals, which are subsequently fed to the FCP module to compute more accurate RTFs and physically-constrained reverberant predictions. 
The updated FCP outputs, together with the direct-path and reverberant signals estimated in the previous iteration, are then provided back to the second DNN, allowing further refinement of the speaker-image estimates. 

\begin{figure}
  \centering
  \includegraphics[width=8.5cm]{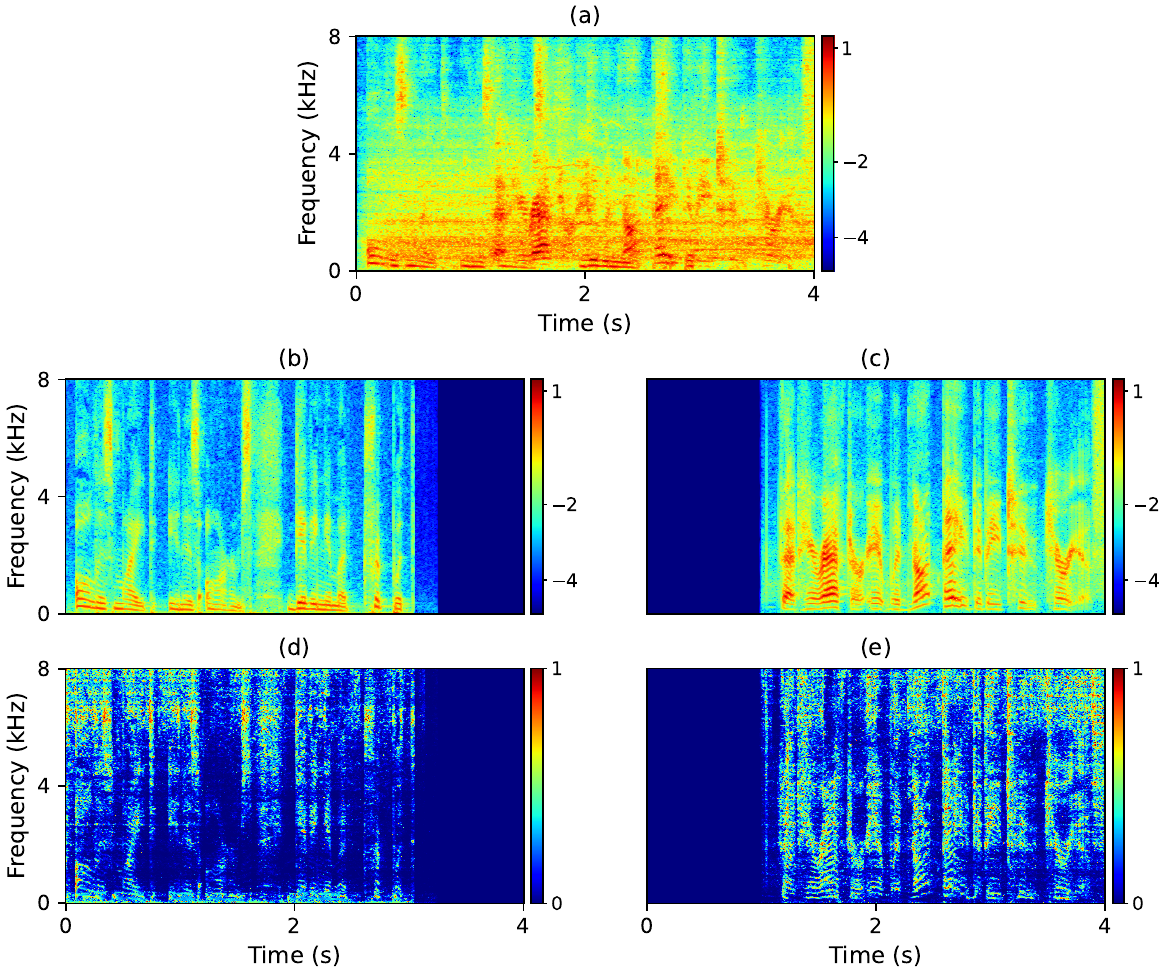}
  \vspace{-0.3cm}
  \caption{
  Illustration, based on a reverberant two-speaker mixture, of log spectrograms of
  (a) mixture, $\text{log}_{10}(|Y|)$;
  (b) direct-path signal of speaker one, $\text{log}_{10}(|S(1)|)$;
  (c) direct-path signal of speaker two, $\text{log}_{10}(|S(2)|)$;
  and ideal ratio masks of
  (d) direct-path signal of speaker one $|S(1)|/(|S(1)|+|Y-S(1)|)$;
  (e) direct-path signal of speaker two $|S(2)|/(|S(2)|+|Y-S(2)|)$.
  See Eq.~(\ref{eq:phy_model}) for the definitions of the symbols.
  }
  \label{figure:fcp_intuition}
  \vspace{-0.6cm}
\end{figure}

Why would this approach work? 
The idea is that, even when room reverberation is strong, there are still many T-F units dominated by the direct-path signal (see the ideal ratio masks plotted in Fig.~\ref{figure:fcp_intuition}(d) and (e)).
Some of these T-F units (e.g., the ones in speech onset) could be easily identified by DNNs since they contain strong direct-path energy and exhibit strong spectro-temporal patterns.
If the RTF can be accurately estimated, based on the identified T-F units (dominated by the direct-path signal), in the subsequent T-F units we could at least reliably figure out the reverberation corresponding to the direct-path signal in the identified T-F units, thereby improving speaker-image separation.

\subsection{Joint Prediction of Direct Signal and Reverberant Speech}

Although the ultimate goal of CxNet is to estimate the reverberant speech, both DNN$_{\text{R\&A,1}}$ and DNN$_{\text{R\&A,2}}$ are designed to jointly predict the direct-path signal and the reverberant speech.
This design is not merely to satisfy the requirement of the FCP module (shown in Fig. \ref{figure:system_overview}), which requires an estimated direct-path signal for generating a physically-constrained estimate of reverberant speech.
More importantly, predicting the direct-path signal provides the network with a cleaner, high-energy, and speaker-specific representation, which exhibits stronger spectro-temporal patterns and can guide the model to better capture the most informative parts of the signal.
Meanwhile, the reverberant signal encodes acoustic context, both of which are essential for realistic and natural-sounding reconstruction.
By jointly learning these complementary aspects, each DNN enforces consistency between the direct-path and reverberant domains, effectively serving as an auxiliary supervision signal that enhances the robustness and accuracy of speaker-image estimation.

\subsection{FCP for RTF Estimation
}\label{FCP_description}

In our system, the direct-path signal $\hat{S}$ is first estimated using a DNN and then used for RTF estimation via FCP \cite{2021_Wang_DEREVERB_FCP_journal, 2021_Wang_DEREVERB_FCP_conference}, thereby enabling speaker-image estimation under explicit physical convolution constraint.
To improve RTF estimation, we extend FCP to a variant with energy-sorted source update.
This section details the two algorithms.

\subsubsection{Adapting FCP for Speaker-Image Separation}%
\label{sec:fcp_review_and_adaptation_for_speaker_image_separation}

Given the DNN-estimated direct-path signal $\hat{S}(c)$, we estimate a $K$-tap, time-invariant FCP filter $\hat{g}(c,f)$ that characterizes the room acoustic response by
\begin{equation}
\hat{g}(c,f) = \arg\min_{g(c, f)}\sum\nolimits_{t}\frac{|Y(t, f)-g(c, f)^\H\tilde{\hat{S}}(c, t, f)|^2}{\hat\lambda (c, t, f)},
\label{eq:FCP_objective}
\end{equation}
where $\widetilde{\hat{S}}(c, t, f) = [\hat{S}(c, t, f), \hat{S}(c, t - 1, f), \dots, \hat{S}(c, t - A + 1, f)]^\T \in \CC^{A}$ stacks a window of current and past T-F units, $(\cdot)^{\H}$ computes Hermitian transpose, and the denominator $\hat\lambda (c, t, f) =\varepsilon\times\max(|Y|^2) + |Y(t, f)|^2$ with $\varepsilon$ flooring the denominators.

Unlike prior FCP applications, which are designed to remove reverberation \cite{2021_Wang_DEREVERB_FCP_journal}, our approach repurposes FCP to preserve reverberation.
The resulting FCP-estimated image
\begin{equation}
\hat{X}^{\text{FCP}}(c, t, f) = \hat{g}(c, f)^\H\tilde{\hat{S}}(c, t, f)
\label{eq:FCP_estimated_image_formation}
\end{equation}
explicitly obeys a physical convolution constraint and can be used as an auxiliary input feature to help the second DNN refine the estimation of speaker images.

\subsubsection{FCP with Energy-Sorted Source Update}%
\label{sec:multi_step_FCP}

\begin{algorithm}[!t]
\footnotesize
\SetKwInOut{Input}{Input}
\SetKwInOut{Output}{Output}
\caption{\footnotesize FCP with energy-sorted source update.}
\label{alg:msFCP_essu}

\Input{Input mixture $Y$ and direct-path estimates $\{\hat{S}(c)\}_{c=1}^C$}
\Output{FCP-estimated speaker images $\{\hat{X}^{\text{FCP}}(c)\}_{c=1}^C$}

Initialize $\hat{X}^{\text{FCP}}(c,t,f)\in \CC$ to zero for each $c$, $t$, and $f$\;
Compute $O = \text{argsort}\Big(\Big[\big\|\hat{S}(c)\big\|_2^2 \text{ for } c=1,\dots,C\Big]\Big)$

\For{$c$ in $O$}{
    Compute $\hat{Z}(c,t,f)=Y(t,f)-\sum_{c',c'\neq c} \hat{X}^{\text{FCP}}(c',t,f)$ 
    Solve $\hat{\mathbf{g}}(c,f)=\underset{\mathbf{g}(c,f)}{{\text{argmin}}} \sum_t \frac{\left|
    \hat{Z}(c,t,f) - \mathbf{g}(c,f)^{\H} \widetilde{\hat{\mathbf{S}}}(c,t,f) \right|^2}{\hat{\eta}(c,t,f)}$, where $\hat{\eta}(c,t,f)= \varepsilon \times \max(|\hat{Z}(c)|^2) + |\hat{Z}(c,t,f)|^2$\;
    Compute FCP-estimated reverberant speaker image for $c$: $\hat{X}^{\text{FCP}}(c,t,f) = \hat{\mathbf{g}}(c,f)^{\H} \widetilde{\hat{\mathbf{S}}}(c,t,f)$\;
}
\end{algorithm}

In multi-speaker scenarios,
the target signal for linear projection used in standard FCP (i.e., the mixture signal $Y$ used in Eq. (\ref{eq:FCP_objective})) may be inaccurate, particularly for weak sources, as the presence of stronger sources in the mixture signal may interfere with the filter estimation. 
By removing the stronger sources beforehand, the FCP estimation for weak speakers can be significantly improved.

Building on this idea, we propose FCP with energy-sorted source update, denoted as FCP-ESSU and detailed in Alg.~\ref{alg:msFCP_essu}, where the FCP filters for different speakers are computed sequentially following an order of descending energy, sorted based on the energy of the estimated direct-path signals (see line $2$ of Alg.~\ref{alg:msFCP_essu}).
For each source $c$, the target signal for linear projection is defined as
\[
\hat{Z}(c) = Y - \sum\nolimits_{c',\, c'\neq c,\|\hat{\mathbf{S}}(c')\|_2^2 > \|\hat{\mathbf{S}}(c)\|_2^2}\hat{X}^{\text{FCP}}(c'),
\]
which removes higher-energy sources before estimating the FCP filters for the weaker ones (see also line $4$ of Alg.~\ref{alg:msFCP_essu}).
We find that this strategy enables more accurate FCP-filter estimation for lower-energy sources, leading to better speaker-image separation.

\section{Experimental Setup}%
\label{sec:experimental_set_up}

This section describes the dataset, DNN configurations, loss function, baseline systems, and miscellaneous system configurations.

\textbf{SMS-WSJ} \cite{2019_Drude_SSP_SMS_WSJ}: 
A benchmark for two-speaker separation in reverberant conditions, provides $33{,}561$ training, $982$ validation, and $1{,}332$ test mixtures at $8$ kHz.
The mixtures are simulated with a $6$-microphone circular array with a diameter of $20$ cm, speaker distances are sample from the range $[1.0, 2.0]$ m, and $\text{T}_{60}$ from $[0.2, 0.5]$ s.
White noise is added at an SNR sampled from the range $[20, 30]$ dB.
We additionally synthesize a three-speaker version of the benchmark using the software provided with SMS-WSJ.

\begin{figure}
\centering
\includegraphics[width=8.5cm]{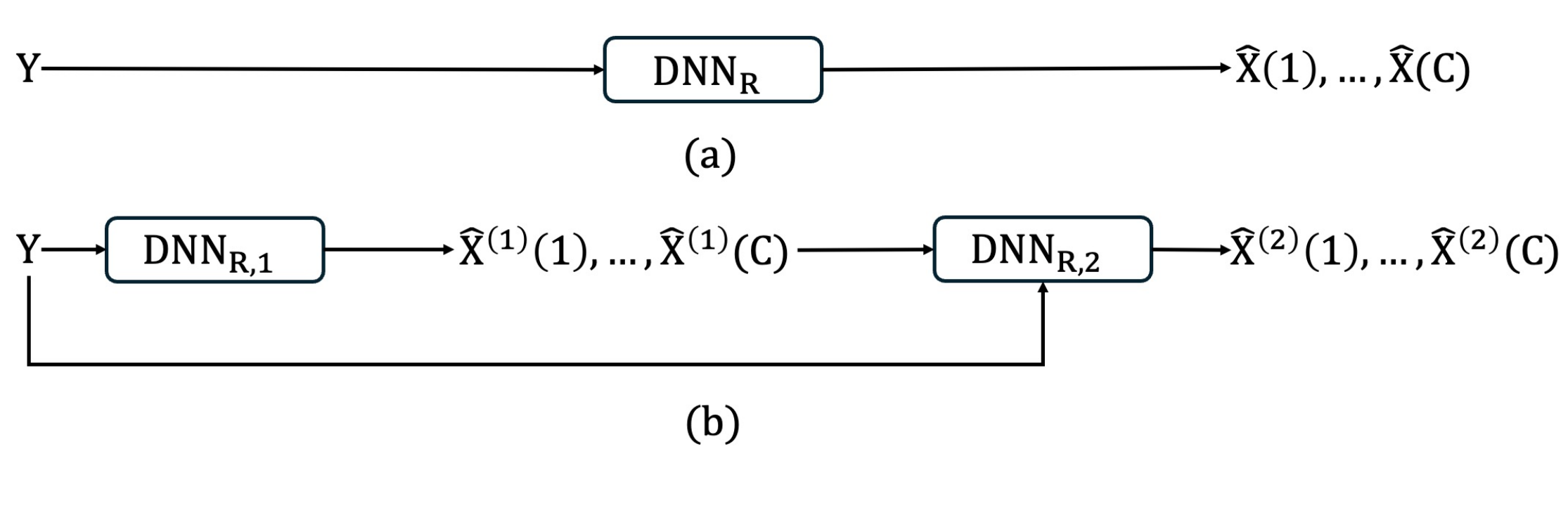}\vspace{-0.6cm}
\caption{
Illustration of 
(a) DNN$_{\text{R}}$; and
(b) DNN$_{\text{R,1}}$+DNN$_{\text{R,2}}$ systems.
}
\label{figure:baseline_system_overview}
\vspace{-0.6cm}
\end{figure}

\textbf{Baseline systems}: We consider two baseline systems, illustrated in Fig.~\ref{figure:baseline_system_overview}.  
System (b), denoted as DNN$_{\text{R,1}}$+DNN$_{\text{R,2}}$, is our proposed framework without additionally estimating the direct-path signal or FCP module.
System (a), DNN$_{\text{R}}$, is a simplified single-DNN variant of system (b).
The size of this single network is chosen to
match the combined size of the two networks in system (b), ensuring a fair comparison in terms of model size.
Notice that we use ``R'' and ``A'' in the subscript (in, e.g., DNN$_{\text{R\&A,1}}$ and DNN$_{\text{R}}$) to denote whether the DNN outputs include \textbf{R}everberant or \textbf{A}nehoic signals.

\textbf{DNN configurations}: We employ TF-GridNet \cite{2022_WANG_SSP_tfgridnet_conference,2022_WANG_SSP_tfgridnet_journal} as the DNN architecture.
Following symbols defined in Table $\mathrm{I}$ of \cite{2022_WANG_SSP_tfgridnet_journal}, we set its hyper-parameters to $D=128, H=200, I=1, J=1$, and $B=6$ blocks for DNN$_\text{R}$ (with $7.7$ M trainable parameters), $B=4$ for DNN$_\text{R\&A,1}$ and DNN$_\text{R,1}$ (both $5.1$ M), and $B=2$ for DNN$_\text{R\&A,2}$ and DNN$_\text{R,2}$ (both $2.6$ M).
DNN$_\text{R\&A}$ jointly predicts direct-path signals and reverberant speaker images.
All DNN modules are trained to perform complex spectral mapping
\cite{2016_Williamson_SSP_complex_ratio_masking,
2017_Fu_SSP_complex_spectrogram_enhancement,
2020_Tan_SSP_complex_spectral_mapping,
2020_Wang_DEREVERB_target_cancellation_complex_spectral_mapping_journal,
2020_Wang_ASR_complex_spectral_mapping,
2021_Wang_PP_distortion,
2021_Wang_SSP_joint_tfgridnet,
2021_Wang_DEREVERB_FCP_journal,
2022_Tan_SSP_neural_spectrospatial_filtering,
2022_WANG_SSP_tfgridnet_journal},
concatenating the real and imaginary (RI) components of input signals to predict the RI components of target signals.
For comparison, we include two external baselines both following system structure in Fig.~\ref{figure:baseline_system_overview}(a): Conv-TasNet \cite{2019_Luo_SSP_single_channel_conv_tasnet} ($5.1$ M) and TF-LocoFormer-M \cite{2024_saijo_SSP_TFLocoformer} ($7.9$ M).

\textbf{Loss functions}:
We leverage three core loss functions to train the DNNs: permutation invariant training (PIT) loss $\mathcal{L}{_\text{PIT}}$ \cite{2017_Kolbak_SS_pit}, mixture-constraint (MC) loss $\mathcal{L}{_\text{MC}}$ \cite{2016_Isik_SSP_deep_clustering}, and enhancement loss $\mathcal{L}{_\text{Enh}}$ \cite{2020_Tan_SSP_complex_spectral_mapping}.
Note that the first DNN module is always trained with $\mathcal{L}{_\text{PIT}}$ to resolve permutation ambiguity, and once it is resolved, the second DNN module is trained in an enhancement fashion.
The loss function for each system is:
(a) \textbf{DNN$_{\text{R}}$}: $\mathcal{L}_{\text{PIT+MC}} = \mathcal{L}_{\text{PIT}} + \mathcal{L}_{\text{MC}}$;
(b) \textbf{DNN$_{\text{R,1}}$}: $\mathcal{L}_{\text{PIT+MC}} = \mathcal{L}_{\text{PIT}} + \mathcal{L}_{\text{MC}}$;
(c) \textbf{DNN$_{\text{R,2}}$}: $\mathcal{L}_{\text{Enh+MC}} = \mathcal{L}_{\text{Enh}} + \mathcal{L}_{\text{MC}}$;
(d) \textbf{DNN$_{\text{R\&A,1}}$}: $\mathcal{L}_{\text{R\&A,1}} = \mathcal{L}_{\text{PIT+MC},R} + \mathcal{L}_{\text{PIT+MC},A}$;
and (e) \textbf{DNN$_{\text{R\&A,2}}$}: $\mathcal{L}_{\text{R\&A,2}} = \mathcal{L}_{\text{Enh+MC},R} + \mathcal{L}_{\text{Enh+MC},A}$.

\textbf{Miscellaneous configurations:} 
For STFT/iSTFT, we use $32$ ms window size, $8$ ms hop size and $256$-point DFT for DNN training,
while FCP uses $128$ ms window, $8$ ms hop, and $1024$-point DFT with filter taps $A$ set to $40$.
Our models are trained and evaluated on the first microphone signal.
The evaluation metrics include SI-SDR \cite{2019_ROUX_SISDR}, narrow-band PESQ \cite{2001_Rix_PESQ} and eSTOI \cite{2016_Jensen_eSTOI} using reverberant speech as the reference signals.

\begin{table}[!t]
\scriptsize
\centering
\sisetup{table-format=2.2,round-mode=places,round-precision=2,table-number-alignment = center,detect-weight=true,detect-inline-weight=math}
\caption{Results of speaker-image separation (2-speaker cases).}
\vspace{-0.3cm}
\label{tab:x_result_table}
\setlength{\tabcolsep}{2pt}
\begin{tabular}{
c
l
S[table-format=1,round-precision=1]
S[table-format=2.1,round-precision=1]
S[table-format=1.2,round-precision=2]
S[table-format=1.3,round-precision=3]
}
\toprule
ID    & {Systems}                                                    & {Iterations} & {SI-SDR(dB)} & {nbPESQ} & {eSTOI} \\
\midrule
{-}   & Unprocessed                                                  & {-}       & -0.034       & 1.868    & 0.603                    \\
\midrule
{1} & DNN$_{\text{R}}$                                             & {-}       & 17.157       & 3.970    & 0.930                    \\
\midrule
{2a} & DNN$_{\text{R,1}}$                                          & {-}       & 15.984       & 3.871    & 0.917                    \\
{2b} & DNN$_{\text{R,1}}$+DNN$_{\text{R,2}}$                       & {1}       & 17.976       & 4.017    & 0.936                    \\
\midrule
{3a} & DNN$_{\text{R\&A,1}}$                                       & {-}       & 17.669       & 3.990    & 0.933                    \\
{3b} & DNN$_{\text{R\&A,1}}$+DNN$_{\text{R\&A,2}}$                 & {1}       & 19.629       & 4.101    & 0.950                    \\
\midrule
{4a} & DNN$_{\text{R\&A,1}}$+FCP+DNN$_{\text{R\&A,2}}$             & {1}       & 20.438       & 4.140    & 0.955                    \\
{4b}& DNN$_{\text{R\&A,1}}$+FCP-ESSU+DNN$_{\text{R\&A,2}}$         & {1}       & 20.777       & 4.152    & 0.958                    \\
{4c}& DNN$_{\text{R\&A,1}}$+FCP-ESSU+DNN$_{\text{R\&A,2}}$         & {2}       & \bfseries 21.390       & \bfseries 4.153    & \bfseries 0.962                    \\
\midrule
{5a}& Conv-TasNet \cite{2019_Luo_SSP_single_channel_conv_tasnet}                  & {-}       & 9.480        & 2.591    & 0.757                    \\
{5b}& TF-LocoFormer-M \cite{2024_saijo_SSP_TFLocoformer}           & {-}       & 16.618       & 3.774    & 0.915                    \\
\bottomrule
\end{tabular}

\vspace{0.1cm}

\scriptsize
\centering
\sisetup{table-format=2.2,round-mode=places,round-precision=2,table-number-alignment = center,detect-weight=true,detect-inline-weight=math}
\caption{Results of speaker-image separation (3-speaker cases).}
\vspace{-0.3cm}
\label{tab:x_result_table_2}
\setlength{\tabcolsep}{2pt}
\begin{tabular}{
c
l
S[table-format=1,round-precision=1]
S[table-format=2.1,round-precision=1]
S[table-format=1.2,round-precision=2]
S[table-format=1.3,round-precision=3]
}
\toprule
ID    & {Systems}                                                    & {Iterations} & {SI-SDR(dB)} & {nbPESQ} & {eSTOI} \\
\midrule
{-}   & Unprocessed                                                  & {-}       & -3.160       & 1.574    & 0.458                    \\
\midrule
{1}  & DNN$_{\text{R}}$                                              & {-}       & 12.853       & 3.498    & 0.859                    \\
\midrule
{2a} & DNN$_{\text{R,1}}$                                            & {-}       & 10.804       & 3.195    & 0.810                    \\
{2b} & DNN$_{\text{R,1}}$+DNN$_{\text{R,2}}$                         & {1}       & 13.175       & 3.496    & 0.858                    \\
\midrule
{3a} & DNN$_{\text{R\&A,1}}$                                         & {-}       & 13.686       & 3.539    & 0.869                    \\
{3b} & DNN$_{\text{R\&A,1}}$+DNN$_{\text{R\&A,2}}$                   & {1}       & 15.567       & 3.759    & 0.901                    \\
\midrule
{4a} & DNN$_{\text{R\&A,1}}$+FCP+DNN$_{\text{R\&A,2}}$               & {1}       & 16.132       & 3.810    & 0.908                    \\
{4b} & DNN$_{\text{R\&A,1}}$+FCP-ESSU+DNN$_{\text{R\&A,2}}$          & {1}       & 16.476       & 3.831    & 0.912                    \\
{4c} & DNN$_{\text{R\&A,1}}$+FCP-ESSU+DNN$_{\text{R\&A,2}}$          & {2}       & \bfseries 17.181       & \bfseries 3.870    & \bfseries 0.921                    \\
\midrule
{5a}& Conv-TasNet \cite{2019_Luo_SSP_single_channel_conv_tasnet}                  & {-}       & 2.201        & 1.722    & 0.499                    \\
{5b}& TF-LocoFormer-M \cite{2024_saijo_SSP_TFLocoformer}           & {-}       & 12.096       & 3.240    & 0.829                    \\
\bottomrule
\end{tabular}
\vspace{-0.6cm}
\end{table}

\section{Evaluation Results}%
\label{evaluation_results}

Table \ref{tab:x_result_table} reports $2$-speaker evaluation results on the SMS-WSJ.

Comparing system $2$a with $2$b, we observe stacking two DNNs producing clear improvements (from $16.0$ to $18.0$ dB).
Comparing system $2$b with $1$, we observe that sequentially training two smaller DNNs (the first one with $4$ TF-GridNet blocks and the second with $2$) outperforms training a larger DNN (with $6$ blocks).

Comparing system $3$a with $2$a, we find that the joint prediction approach produces clear improvement ($17.7$ vs. $16.0$ dB SI-SDR).
The improvement is attributed to
jointly predicting direct-path signal and reverberant speech, which enables more accurate reconstruction of the reverberant speech.
Stacking one DNN in $3$b produces further gains over $3$a, in consistent with the trend in 2a and 2b.

In system $4$a, we insert the FCP module described in Section \ref{sec:fcp_review_and_adaptation_for_speaker_image_separation} in between DNN$_{\text{R\&A,1}}$ and DNN$_{\text{R\&A,2}}$.
This leads to clear improvement over 3b ($20.4$ vs. $19.6$ dB SI-SDR), which indicates the effectiveness of the proposed neural forwarding filtering approach for speaker-image separation.
We enhance the system in Fig.~\ref{figure:system_overview} by replacing FCP with FCP-ESSU described in Section \ref{sec:multi_step_FCP}.
As shown in $4$b, this change improves SI-SDR from $20.4$ to $20.8$ dB, demonstrating the effectiveness of the ESSU strategy in multi-speaker scenarios.
Our final CxNet system in Figure~\ref{figure:system_overview} adopts FCP-ESSU. 
From system $4$b and $4$c, we observe that further iterating DNN$_2$ at run time can enhance performance.
With $2$ iterations of DNN$_2$, the system in $4$c achieves an improvement of $3.4$ dB SI-SDR, $0.13$ nbPESQ, and $0.026$ eSTOI over the strongest baseline, system $2$b, which does not exploit the physical convolution constraint.

In system $5$a and $5$b, we provide the results of Conv-TasNet and TF-LocoFormer-M.
Their performance is clearly worse than our best-performing system.

In Table \ref{tab:x_result_table_2}, we report the results on three-speaker-image separation. Similar trend as in the two-speaker case is observed.

\begin{figure}
\centering
\includegraphics[width=8.5cm]{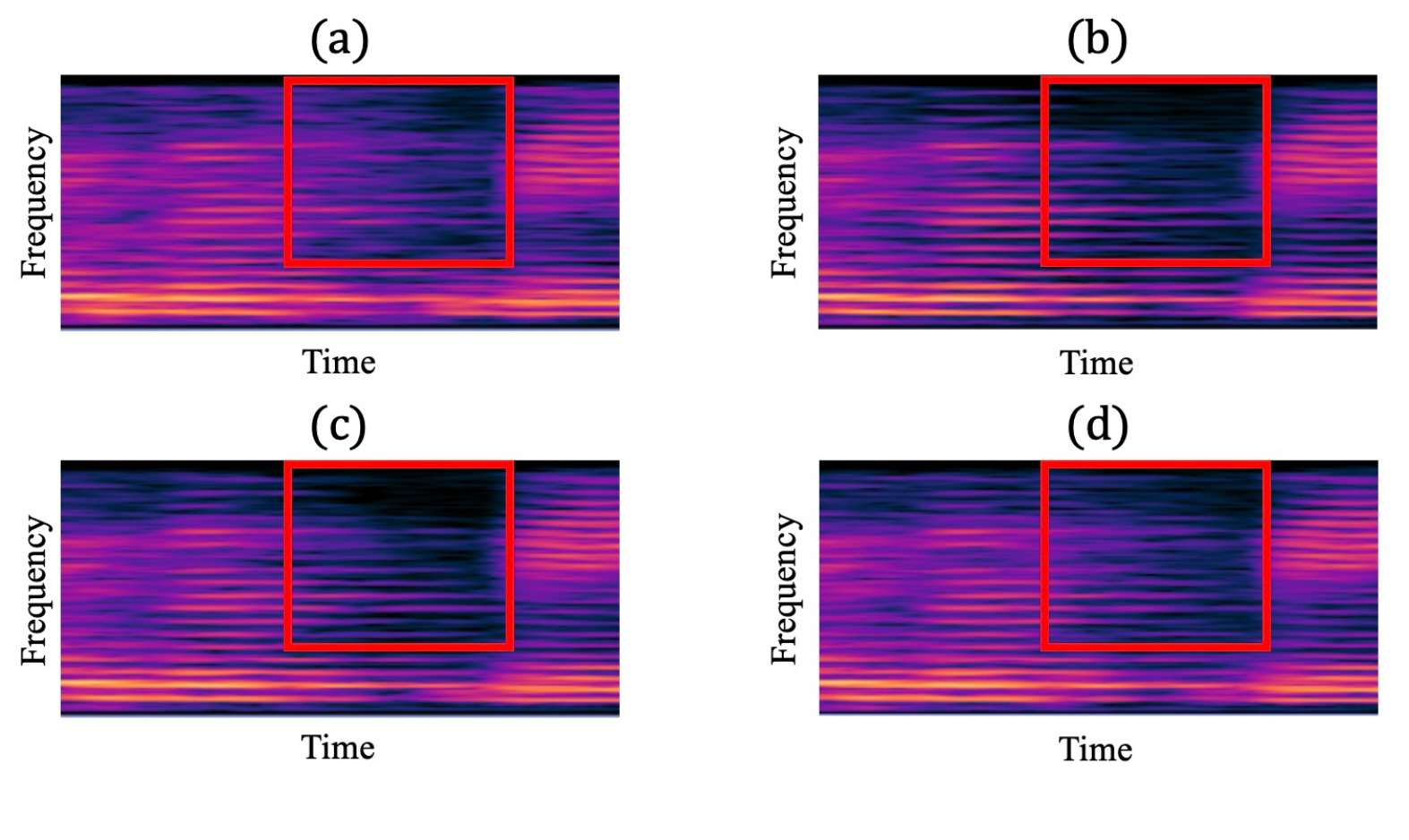}\vspace{-0.6cm}
\caption{
Output spectrograms of system (b) $2$b; (c) $3$b; and (d) $4$b on an SMS-WSJ mixture, along with (a) ground-truth spectrogram. 
}
\label{figure:late_reverb_observation}
\vspace{-0.3cm}
\end{figure}

\begin{figure}
\centering
\includegraphics[width=8.5cm]{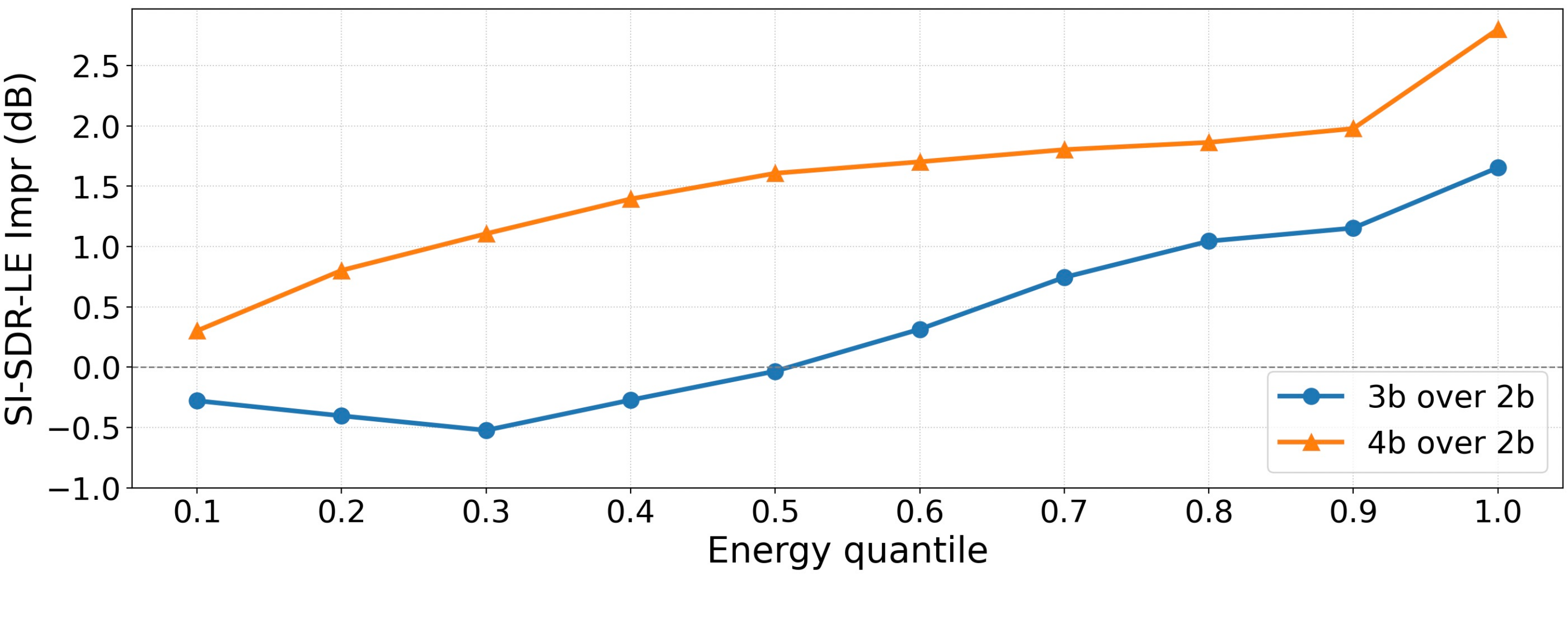}\vspace{-0.6cm}
\caption{
SI-SDR-LE improvements of system $3$b and $4$b over $2$b at different energy quantiles.
}
\label{figure:lhsisdr}
\vspace{-0.5cm}
\end{figure}

We further investigate the proposed system through
qualitative and quantitative analysis of system $2$b, $3$b, and $4$b in Fig.~\ref{figure:late_reverb_observation} and \ref{figure:lhsisdr}, targeting at understanding how the joint prediction framework and the incorporation of FCP
improve speaker-image estimation.

In Fig.~\ref{figure:late_reverb_observation}, we shows example outputs from the three systems on a mixture sampled from SMS-WSJ, where the red box highlights a low-energy late reverberation region for comparison.
In Fig.~\ref{figure:late_reverb_observation}(d), we observe that CxNet with FCP (i.e., system $4$b) better recovers late reverberation than (b) and (c) (corresponding to system $3$b and $2$b), which do not explicitly leverage FCP modeling.

So far, all the evaluation scores are computed based on the entire separated signal. However, this does not reflect the performance of different algorithms at T-F units where the target speech has low energy, such as at the T-F units only containing late reverberation.
To address this, based on the true target reverberant speech we propose to first compute a binary T-F mask, which is set to $0$ at a T-F unit if its energy is larger than a pre-defined energy threshold and to $1$ otherwise, and then use this mask to mask the estimated and true reverberant speech and compute SI-SDR.
We name this metric \texttt{SI-SDR-LE}, where ``LE'' means low-energy T-F units.
In Fig.~\ref{figure:lhsisdr}, we quantitatively show the SI-SDR-LE improvements of systems $3$b and $4$b over $2$b, by setting the pre-defined energy threshold to an energy quantile computed based on the energy of the T-F units of the true target reverberant speech.
From the \texttt{3b$\,$over$\,$2b} curve, we observe that including direct-path prediction steers the model towards stronger-energy T-F units, yielding clear gains for quantiles above $0.5$, indicating that joint prediction can improve the model’s separation ability at higher-energy T-F units.
However, its effectiveness diminishes in lower-energy T-F units (quantiles below $0.5$), where the improvement turns negative.
In comparison, the \texttt{4b$\,$over$\,$2b} curve shows consistently positive improvement across all energy quantiles, clearly outperforming \texttt{3b$\,$over$\,$2b}.
This indicates that including the FCP module can mitigate the limitation of joint prediction and boost the estimation in lower-energy T-F units while further improving the performance in stronger-energy T-F units.

\vspace{-0.2cm}
\section{Conclusions}\label{sec:conclusion}
\vspace{-0.2cm}

We present CxNet, a novel neural architecture for speaker-image separation that employs neural forward filtering for enhanced performance.
CxNet jointly predicts direct-path signal and reverberant speech for each speaker, using the cleaner, more informative direct-path representation to guide estimation of reverberant output.
The system incorporates a forward convolutive prediction (FCP) module, explicitly modeling the linear convolution between each speaker’s direct-path signal and reverberant image, providing physically consistent features that improve estimation accuracy.
We also introduce an energy-sorted variant, FCP-ESSU, which further improves performance by reducing the influence of stronger sources when estimating weaker ones.
Experimental results on the SMS-WSJ dataset show clear improvement over baselines for both two- and three-speaker mixtures, while maintaining comparable model complexity.

\newpage
\bibliographystyle{IEEEtran}
{\footnotesize
\bibliography{references.bib}
}
\end{document}